GaAs:Mn nanowires grown by molecular beam epitaxy of (Ga,Mn)As at MnAs segregation conditions

*J. Sadowski[1,2,3], P. Dluzewski[2], S. Kret[2], E. Janik[2], E. Lusakowska[2], J. Kanski[4], A. Presz[5], F. Terki[1], S. Charar[1], D. Tang[6]

(1) Groupe d'Etude des Semiconducteurs, Universite de Montpellier 2, Montpellier, France
(2) Institute of Physics, Polish Academy of Sciences, Warszawa, Poland
(3) MAX-Lab, Lund University, Sweden
(4) Department of Applied Physics, Chalmers University of Technology, Goteborg, Sweden
(5) Institute of High Pressure Physics (UNIPRESS), Polish Academy of Sciences, Warszawa, Poland
(6) FEI Co, NL-5600 MD Eindhoven, Netherlands

ABSTRACT:

GaAs:Mn nanowires were obtained on GaAs(001) and GaAs(111)B substrates by molecular beam epitaxial growth of (Ga,Mn)As at conditions leading to MnAs phase separation. Their density is proportional to the density of catalyzing MnAs nanoislands, which can be controlled by the Mn flux and/or the substrate temperature. After deposition corresponding to a 200 nm thick (Ga,Mn)As layer the nanowires are around 700 nm long. Their shapes are tapered, with typical diameters around 30 nm at the base and 7 nm at the tip. The wires grow along the <111> direction, i.e. along the surface normal on GaAs(111)B and inclined on GaAs(001). In the latter case they tend to form branches. Being rooted in the ferromagnetic semiconductor (Ga,Mn)As, the nanowires combine one-dimensional properties with the magnetic properties of (Ga,Mn)As and provide natural, self assembled structures for nanospintronics.

* corresponding author, e-mail: janusz.sadowski@maxlab.lu.se

Self assembled one-dimensional structures such as nanowires (NW) or nano-whiskers are currently subject of intense research activity.[1,2] They are expected to possess great capabilities for a wide range of applications, ranging from physical nano-devices such as diodes,[1] transistors in both classical and single electron operation mode,[3] and nano-sensors,[4,5] to biological nano-engineering tools.[6] In view of these perspectives it is highly desirable to master the ways of NW formation and to understand physical phenomena leading to self assembled nanowires growth. Although the basic mechanism leading to the growth of nanowire structures was identified over four decades ago by Wagner and Ellis[7] as so called vapor-liquid solid (VLS) growth, nowadays several other mechanisms have been proposed, and their applicability is debated.[8,9] The "classical" VLS mechanism of NWs formation requires the presence of liquid nano-droplets of a catalyst material, which accommodates and dissolves the NW constituent elements, and delivers them, due to supersaturation, to the NW tip at the NW-catalyst interface. This mechanism has been used for successful description of NW growth in different materials systems, both single elements (Si, Ge) and multinary alloys (e.g. binary and ternary III-V and II-VI semiconductors).[1,10-12] However, there is quite a large variety of materials that do not need any catalyst for NW growth, e.g. GaN[13] or ZnO[14]. There are also reports that the catalyst nanoparticles at the NW tips do not necessarily need to be in the liquid phase to stimulate the NW growth.[8,15] Here we present an example of a material system, (Ga,Mn)As in which the NW formation is rather unexpected, but as it occurs, it is particularly interesting in view of opening prospects to integrate two important research domains, spintronics and self-assembled nanostructures. In the (Ga,Mn)As ternary alloy a significant part of the Ga atoms (up to about 8 at%) is replaced by Mn. The Mn atoms provide both the localized spins $S=5/2$ and charge carriers (valence band holes), since they act as shallow acceptors in the GaAs host. The coupling between spins of localized Mn ions and valence band holes leads to a ferromagnetic phase in (Ga,Mn)As for Mn content higher than about 0.5 at% and sufficiently high concentration of valence band holes (in the range of $10^{20}$ cm$^{-3}$). (Ga,Mn)As is nowadays considered as a prototype ferromagnetic semiconductor.[16] Since the equilibrium solubility of Mn in GaAs is below 0.1at%, (Ga,Mn)As with the required Mn concentrations in the range of several at% can only be obtained by a highly non-equilibrium growth method such as low temperature molecular beam epitaxy (MBE). The typical MBE growth temperatures used for GaAs (590 – 640 °C) are not applicable for (Ga,Mn)As, because under high temperature growth conditions the Mn delivered into the growing GaAs film segregates to MnAs nano-clusters.[17] As shown for the first time by Ohno et al.,[18] the use of very low growth temperatures, in the range of 200 – 300 °C allows overcoming the Mn segregation problems and leads to formation of a (Ga,Mn)As ternary alloy. (Ga,Mn)As has been already demonstrated to be useful for fabrication of functioning spintronics components such as spin lasers,[19] or tunneling magnetoresistance structures.[20] There are reports on self assembled NWs obtained in other magnetic semiconductor systems like (Ga,Mn)N, (Zn,Mn)O.[21,22] However, in these materials the realization of carrier-induced ferromagnetism is uncertain due to extreme difficulties in sufficiently high p-type doping levels, and evidence of segregation of additional magnetic phases identified in the corresponding layer structures.[23,24] In the case of (Ga,Mn)As the carrier induced ferromagnetism is very well established and quite well understood, at least in the layer geometries.[25] Recently magnetoelectronic effects associated with single electron transport were reported in (Ga,Mn)As nanostructures obtained via chemical methods (e-beam lithography and wet chemical etching).[26] Thus the possibility to obtain self assembled wire-like nanostructurers in this compound is particularly interesting. We note that two reports concerning Mn induced growth of GaAs nanowires have appeared recently. Martelli et al.[27] describe Mn-catalyzed formation of GaAs nanowires during GaAs MBE growth at moderate and high temperatures (450 to 620 °C). These authors deposited a thin metallic Mn layer prior to the NW growth and observed catalytic properties of pure Mn, revealing its presence at the NW tips. Jeon et. al.[28] ascribe the development of defect-free (Ga,Mn)As nanowires (containing 20% Mn) to initial formation on (Ga,Mn)As islands due to lattice mismatch between the high concentration (Ga,Mn)As layer and the GaAs(001) substrate. However, as the lattice mismatch between (Ga,Mn)As and GaAs is quite small – about 0.4% for the highest Mn content possible (close to 10%), and even for 1000 nm thick (Ga,Mn)As layers deposited on GaAs(001) substrates the layer is coherently strained to the substrate without any lattice relaxation,[29] this mechanism is questionable. In this letter we show that the mechanism leading to the development of nanowires in this system is the catalyzing action of MnAs nano-dots appearing due to phase separation occurring during (Ga,Mn)As MBE growth at the temperature which is too high for the formation of a uniform (Ga,Mn)As ternary alloy.

The samples were grown in a KRYOVAK MBE system dedicated to (Ga,Mn)As growth. As substrates we used epi-ready GaAs(001) and GaAs(111)B n-type wafers. To facilitate direct comparison, pieces of the two substrates were placed close to each other on the same substrate holder. The substrate temperature during growth was monitored by an infrared pyrometer operating in the 100 – 700 $^{o}$C temperature range. After growing a thin GaAs buffer layer at high temperature (590 oC), the substrate temperature (TS) was decreased to 300 - 350 $^{o}$C (depending on Mn flux - higher TS for lower Mn flux) and (Ga,Mn)As growth was started. The growth was performed using As2 molecular flux generated by a valved cracker As source, with a Ga:As flux ratio of about 2. The Mn flux was evaluated via RHEED intensity oscillations recorded for (Ga,Mn)As thin film calibration samples. The substrate temperature used for NW growth was set to be above the maximum temperature growth of uniform (Ga,Mn)As layers.[17,30] MBE growth at these conditions leads to the appearance of 3D-features in RHEED, due to formation of MnAs clusters. In the case of (Ga,Mn)As with high Mn content (in the range of 5%) growth above a critical temperature results in an abrupt transition between 2D and 3D RHEED features. The morphology of a (Ga,Mn)As surface with MnAs clusters investigated by AFM is shown in Fig. 1. The MnAs islands have diameters of 300-500 Å and heights in the range 20 – 50 Å.

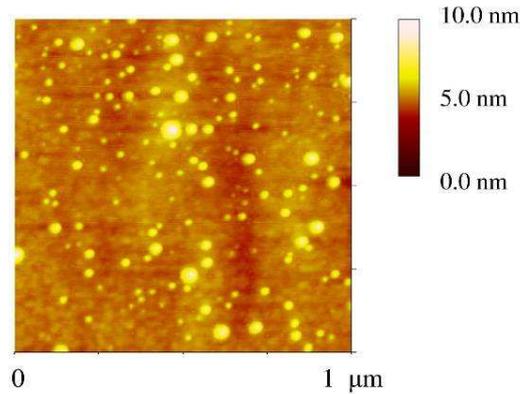

*Fig. 1. AFM image of a 50 nm thick $Ga_{0.94}Mn_{0.06}As$ layer with MnAs nano-islands segregated at the surface.*

For low Mn content (1% in the present case) the transition from 2D to 3D surface is rather gradual, as shown in Figs. 2 and 3. For both (001) and (111)B substrates the growth starts from smooth two-dimensional GaAs surfaces. The initial (001) surface (Fig. 2a) has a two-fold reconstruction in [110] and [-110] azimuths and four-fold in [010] and [100], typical for the c(4x4) reconstructed GaAs(001) surface. After opening Mn and Ga shutters the two-fold periodicity in the [110] azimuth disappears and surface roughening occurs, as manifested by the 3D chevron-like features at the main diffraction streaks. After about 5 minutes of the growth the 2D diffraction streaks disappear and a purely 3D diffraction image is seen. The multiple diffraction spots are due to coherent twinning along <111> direction. Fig. 2g shows a scanning electron microscopy (SEM) image of the NWs grown on (001) substrate after 1 hour growth time. The SEM images were obtained with the incident electron beam inclined 45° with respect to the surface plane. The nanowires have different orientations and many of them have branches, pointing preferentially away from the substrate surface.

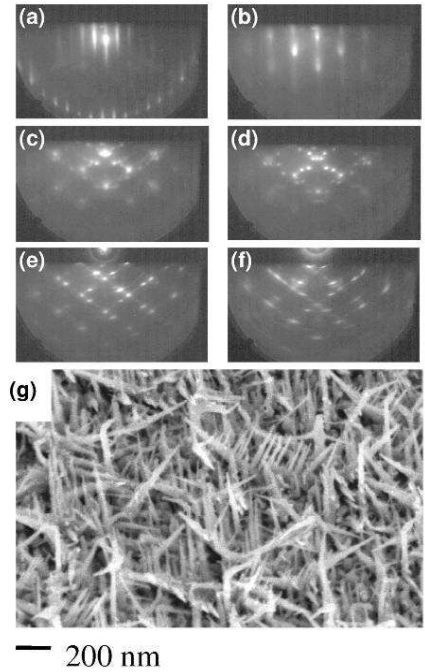

*Fig. 2. RHEED images along [110] azimuth at different stages of growth on GaAs(001). (a) - initial GaAs(001) surface with two-fold reconstruction, (b)-(f) after 3, 6, 11, 25, and 60 min of growth; and a SEM picture after 60 min growth - (g).*

The evolution of the RHEED patterns for nanowires grown on the GaAs(111)B substrate is different than for the (001) substrate. The RHEED images (Fig3 a-f) indicate much less disordered structure. This is consistent with the SEM image (Fig. 3g), which shows that the majority of nanowires are now perpendicular to the substrate surface (i.e. oriented along the <111> growth direction). In this case there are no additional spots between substrate diffraction streaks visible in RHEED. All spotty patterns are placed along the 2D RHEED streaks from the initially smooth GaAs(111)B surface, since the twinning plane is now parallel to the GaAs(111)B substrate plane.

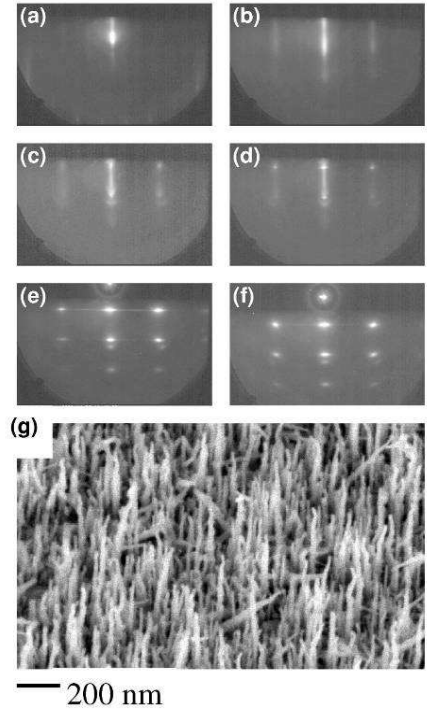

*Fig. 3. RHEED images along the [11-2] azimuth at different stages of growth on GaAs(111)B. (a) - initial GaAs(111)B surface, (b)-(f) after 2, 7, 10, 25, and 60 min of growth; and a SEM picture after 60 min growth - (g).*

Transmission electron microscopy (TEM) images from individual nanowires scratched out off the (001) substrate and placed on a carbon film TEM grid are shown in Fig. 4a. All the nanowires are tapered. The single nanowire shown in Fig. 4b is quite smooth on one side and much rougher on the other. As noted above, the branches visible in SEM (Fig. 2g) are found preferentially on one side of the NW. The NW morphology is better visible in a TEM picture made in cross-sectional geometry, where both NW and substrate surface are visible. Fig. 4c shows the cross-sectional TEM picture of a sample grown for 1h with Mn content corresponding to 1at% in a uniform growth of (Ga,Mn)As layer. It is clear that the smooth NW side is facing the substrate, while the short branches occur on the "outer" side, i.e. the side facing the effusion sources. It seems likely that the roughness seen on one NW side in Fig 4b reflects an early stage of branch formation. Their geometrical distribution indicates that the branches are formed due to the action of impinging Mn atoms, which are known to have very low surface mobilities, especially at the low growth temperatures used. It is worth to notice that the density of branches increases with increasing Mn flux.

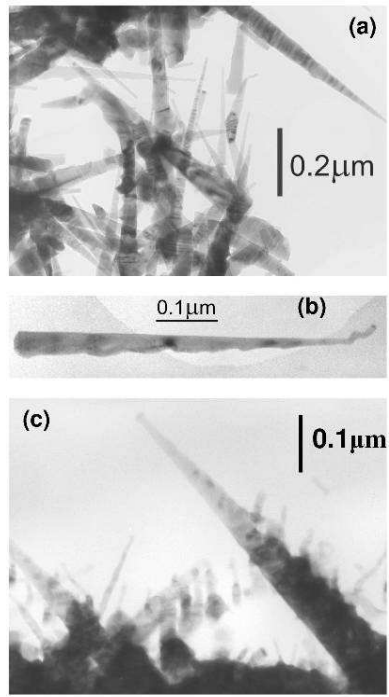

*Fig. 4. TEM images: (a) GaAs:Mn NWs scraped from the (001) substrate onto a holey carbon film; (b) a single NW revealing its straight and rough sides. (c) cross-sectional view along the <110> substrate direction. The downward NW sides are straight without secondary branches, contrary to the sides exposed to the effusion sources.*

Fig 5 shows a result of an EDX measurement of a chemical composition along a top part of a selected nanowire. A clear enhancement of Mn and As signals at a tip is found. Because the Mn:As ratio at the tip is close to one, and no free Mn can occur in the III-V MBE system with high As overpressure, Mn at the tip region must be bound with As forming a MnAs nano-cluster.

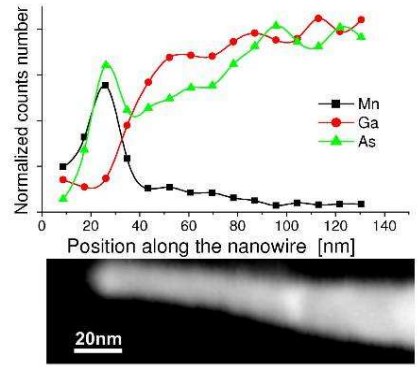

*Fig. 5. EDX line scan along the top part of the GaAs:Mn nanowire shown below. The presence of MnAs at the nanowire tip is manifested by the enhanced Mn and As contents.*

The magnetic properties of one of the GaAs:Mn NW samples were analyzed by a SQUID magnetometer, see Figure 6. The sample was grown for 60 min on GaAs(100) at 320 °C, with a Mn flux corresponding to 6% Mn content in uniform (Ga,Mn)As and at a planar growth rate of 0.2 monolayer/s. Because the SQUID signal is collected from all parts of the sample (including GaMnAs layer, nanowires, and quantum dots), one can expect rather complex overall magnetic characteristics. Nevertheless, some significant features are observed. First, we note that the sample shows nonzero magnetization and hysteretic behavior up to 170 K (M(H) curves are displayed only up to 90 K). Second, we see that the magnetization curve (only field cooled data were recorded) has a clear break around 30 K. These observations can be correlated with results reported for Mn(Ga)As nanoclusters obtained by thermal annaling of (Ga,Mn)As layers.[31] Specifically, it was found that slow annealing in the growth chamber resulted in MnAs clusters in ZB structure, and clusters subjected to long term annealing at temperatures similar to those applied in the present work showed no room-temperature ferromagnetism. Instead, the samples exhibited superparamagnetic behavior with a blocking temperature around 30 K. Although further detailed magnetic characterization of the present samples is needed, we tentatively associate the ferromagnetic signal up to 170 K with the (GaMn)As layer formed parallel with the nanowires, and the rapid increase in the magnetization below 30 K with the nanoclusters located on the nanowires.

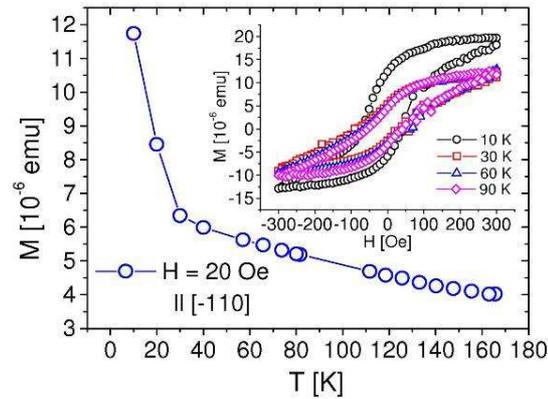

*Fig. 6. Temperature dependence of magnetization (main plot) and hysteresis curves (inset) measured at 10, 30, 60, and 90 K for nanowires grown on GaAs(001) for 1h at 320 °C substrate temperature, with Mn flux corresponding to 6% Mn content in uniform (Ga,Mn)As and at a planar growth rate of 0.2 ML/s. M(T) dependence was measured with external magnetic field of 20 Oe oriented along the [-110] direction of GaAs(001) substrate.*

In conclusion, we have observed formation of nanowires during MBE growth of (Ga,Mn)As at MnAs segregation conditions. Although the nanowires appear without any external catalyst, the NW growth is induced by surface segregated MnAs nanoclusters. The GaAs:Mn nanowires are tapered and grow along the <111> direction both on (001) an (111)B oriented GaAs substrate. NWs grown on the GaAs(001) substrate have a strong tendency to form branches at the sides exposed to the fluxes from effusion cells during the MBE growth.

ACKNOWLEDGMENTS.

The research was partially supported by the Ministry of Science and Higher Education (Poland) through grants: N202-052-32/1189, N507 030 31/0735 and by the Network "New materials and sensors for optoelectronics, information technology, energetics and medicine". J.S. acknowledges support by the research project financed by EADS (France). The (Ga,Mn)As project in Lund is supported by grants from the Swedish Research Council (VR).